\documentstyle[12pt]{article}
\makeatletter
\@addtoreset{equation}{section}
\makeatother

\begin{document}
%
 \mbox{} \hspace{1.5cm}July  1992 \hspace{7.4cm}HLRZ-92-51\\
\begin{center}
\vspace*{0.5cm}
{{\large Chern-Simons term in the  \\
4-dimensional SU(2) Higgs Model}
 } \\
\vspace*{0.7cm}
{\large F.~Karsch$^{1,2}$,
        M.~L.~Laursen$^{1,}
$\footnote{Talk presented by M.~L.~Laursen at the HLRZ workshop
on Dynamics of First Order Phase Transitions, June 1-3, 1992.}  \\
        T.~Neuhaus$^2$, B.~Plache$^2$ \\
        and \\
        U.~-J.~Wiese$^{3,}
        $\footnote{Supported by the Schweitzer Nationalfond}}\\
\vspace*{0.7cm}
{\normalsize
$\mbox{}^1$ {HLRZ, c/o KFA J\"{u}lich,
             P.O. Box 1913, D-5170 J\"{u}lich, Germany}\\
$\mbox{}^2$ {Fak. f. Physik, Univ. Bielefeld,
             D-4800 Bielefeld, Germany}\\
$\mbox{}^3$ {Inst. f. Theor. Physik, Univ. Bern,
             CH-3012 Bern, Switzerland}}\\
\vspace*{2cm}
{\bf Abstract}
\end{center}
\setlength{\baselineskip}{1.3\baselineskip}

Using a variation of Lueschers geometric charge definition for SU(2)
lattice gauge theory, we have managed to give a geometric
expression for it's Chern-Simons term.
{}From this definition we have checked the periodic structure. We
determined the Chern-Simons density for lattices $L^4$
and $L^3 \times 2,\, 4$
with $L=4,\,  6,\mbox{ and }8$
near the critical region in the SU(2) Higgs model. The data indicate
that tunneling is increased  at high temperature.
\newpage

\section{Introduction}
Quite some time ago 't Hooft found that the baryon number and the
lepton number are not conserved in the electroweak theory \cite{Hooft}
$B-L$ is of course conserved due to the anomaly cancellation.
Baryon number violation is
caused by the nontrivial topological winding of the SU(2) gauge fields.
The anomaly of the fermionic current relates the winding of the gauge
fields and the change in the baryon number by an amount
\begin{equation}
 B(t_2) - B(t_1) = \frac{N_f}{16\pi^{2}}
                   \int_{t_1}^{t_2} \int d^{3}x
                    tr[F_{\mu\nu}\tilde{F}_{\mu\nu}]
\end{equation}
where $N_f$ is the number of families of quarks and leptons.
Equivalent, in the temporal gauge $A_0=0$
we can relate the change in the baryon number
to the change in the Chern-Simons number
\begin{equation}
            B(t_2)-B(t_1) = N_{f}[N_{CS}(t_2)-N_{CS}(t_1)]
\end{equation}
where the Chern-Simons number  $N_{CS}$ is
\begin{equation}
           N_{CS}= - \frac{1}{8\pi^{2}} \int d^{3}x
   \epsilon_{ijk} tr[A_{i}(\partial_{j}A_{k}+\frac{2}{3}A_{j}A_{k})].
\end{equation}
At zero temperature such processes are exponentially suppressed as
$exp(-2\pi/\alpha_W)$, $\alpha \approx 1/30$. This is because any
gauge field configuration which changes the winding number has an action
at least that of the barrier height
$2\pi/\alpha_W$.

At high temperatures which prevail in the early
universe such an exponential suppression is absent since the system can
pass over the barrier classically. The only suppression factor is the
Boltzmann factor $exp(-\beta E)$ where $E$ is the barrier height, and it
was shown by the authors in
ref.~\cite{Kuzmin} that this factor goes to one.
Thus, any baryon asymmetry generated at the GUT scale will
get washed out as the universe approaches
the electroweak phase transition
from above. This happens typically around 10 Tev and is caused by
static objects called sphalerons. These are unstable,
but finite energy solutions of the classical Yang-Mills Higgs fields.
While an instanton  tunnel between
two vacua, a sphaleron  moves from one top of the barrier to the next.
If we assume that the vacuum has $N_{CS}=0$, then the sphaleron has
a baryonic charge of 1/2.

The radical solution to the
problem of the observed baryon asymmetry in the universe was put forward
by the same authors using CP violating processes in the electroweak
theory and assuming thermal non-equilibrium provided by the expansion
of the universe. In order for the whole scenario to work, it is necessary
that the electroweak phase transition is of first order, which severely
restricts the Higgs and the top quark masses. For more on the origin
of the baryon asymmetry see ref.~\cite{Kolb}.

There has been some lattice studies of baryon number violating
processes in the two dimensional Abelian Higgs model, as well as in the
four dimensional SU(2) Higgs model
ref.~\cite{Grig} and ref.~\cite{Ambjorn}. The configurations
are prepared at high temperature and the system is allowed either
to change via the classical Hamiltonian equation of motion or
by a Langevin equation. It is then possible to study how the
Chern-Simons term $N_{CS}$ changes during the evolution
as a function of the temperature. It was shown that when
$\Delta N_{CS} = \pm 1$ the system passes through a sphaleron transition.

Since  baryon number violating processes are of
non-perturbative origin it is not clear if the sphaleron
approximation is sufficient enough. There might be other relevant
configurations contributing. This warrants a study in the full Higgs
model without assuming high temperature. The quantity we have studied is
the constraint effective potential $V(N_{CS})$
\begin{eqnarray}
\exp(-V(N_{CS})) &  = & \int dAd\phi \exp(-S(A,\phi)) \nonumber \\
                 &    & \delta (N_{CS}+ \frac{1}{8\pi^{2}} \int d^{3}x
   \epsilon_{ijk} tr[A_{i}(\partial_{j}A_{k}+\frac{2}{3}A_{j}A_{k})]).
\end{eqnarray}
As      a measure for baryon number violating processes we can use
the parameter $r=\exp(V(0)-V(\frac{1}{2}))$.
To evaluate $N_{CS}$ we     have
used a geometric definition given in
ref.~\cite{Luesch}. After vectorizing
the code first made by one of us (MLL) , see ref.~\cite{Fox},
we are able to calculate $N_{CS}$ efficiently.

\section{Topological charge and the Chern-Simons term in the continuum}
As a warm up, let us first consider the two dimensional Abelian gauge
theory. We always work in Euclidean space time.
The theory is determined from the gauge potentials $A_{\mu}(x)$,
($\mu = 0,1$)
which under a local gauge transformation $g(x)$ transform as:
\begin{equation}
 A_{\mu}(x) \rightarrow  A_{\mu}(x) + ig^{-1}(x)\partial_{\mu}g(x).
\end{equation}
The gauge field tensor $F_{\mu\nu}(x)$:
\begin{equation}
 F_{\mu\nu}(x) =
      \partial_{\mu}A_{\nu}(x) - \partial_{\nu}A_{\mu}(x)
\end{equation}
is of course gauge invariant. From this we can define a gauge invariant
and integer topological charge $Q$:
\begin{equation}
 Q = \frac{1}{4\pi} \int_{M} d^{2}x \epsilon_{\mu\nu} F_{\mu\nu}(x)
 \in Z
\end{equation}
with $M$ being the manifold. The topological charge density $q(x)$ can
be written as a perfect derivative
\begin{equation}
 q(x)   =   \frac{1}{4\pi} \epsilon_{\mu\nu} F_{\mu\nu}(x)
        =  \partial_{\mu}K_{\mu}(x)
\end{equation}
where we have introduced the Chern-Simons density $K_{\mu}(x)$.
\begin{equation}
 K_{\mu} = \frac{1}{2\pi} \epsilon_{\mu\nu} A_{\nu}(x).
\end{equation}
It transforms like:
\begin{equation}
 K_{\mu}(x) \rightarrow K_{\mu}(x) + \frac{i}{2\pi}
  \epsilon_{\mu\nu} g^{-1}(x)\partial_{\nu}g(x).
\end{equation}
Now we can define the gauge dependent Chern-Simons number $N_{CS}$
\begin{equation}
 N_{CS} = \int_{\partial M = S^1} dx_{1} K_{0}(x) \not \in Z
\end{equation}
Here, the boundary of the manifold $\partial M$
is assumed to be a one sphere $S^1$.
In general $N_{CS}$ is not an integer. This
happens only if the the gauge field is pure gauge, that is
$A_{\mu}(x) = ig^{-1}(x)\partial_{\mu}g(x)$. However the gauge variation
is always an integer
\begin{equation}
 \delta N_{CS} = \frac{i}{2\pi} \int_{\partial M=S^1} dx_{1}
                    g^{-1}(x)\partial_{1}g(x) \in Z.
\end{equation}
This follows from homotopy theory by considering the mapping
\begin{equation}
  g(x): S^{1} \rightarrow U(1) = S^{1}.
\end{equation}
Such mappings are characterized with the homotopy class
\begin{equation}
\Pi_{1}(S^{1}) \in Z.
\end{equation}

Now we are prepared for the four dimensional SU(2)   gauge theory.
The gauge fields are now non-abelian
$A_{\mu}(x) = A_{\mu}^{a}(x)T^{a}$
($\mu = 0,1,2,3$ and  $T^{a}, a=1,2,3 $ are the Pauli-matrices),
and they transform
under a local gauge transformation $g(x)$  as:
\begin{equation}
 A_{\mu}(x) \rightarrow
 g^{-1}(x)A_{\mu}(x)g(x) + g^{-1}(x)\partial_{\mu}g(x).
\end{equation}
The gauge field tensor
$F_{\mu\nu}(x)$:
\begin{equation}
 F_{\mu\nu}(x) = \partial_{\mu}A_{\nu}(x) - \partial_{\nu}A_{\mu}(x)
      + [A_{\mu}(x),A_{\nu}(x)]
\end{equation}
transforms   now gauge covariantly
\begin{equation}
 F_{\mu\nu}(x) \rightarrow
 g^{-1}(x)F_{\mu\nu}(x)g(x)
\end{equation}
In analogy with two dimensions we have a gauge invariant and integer
topological charge $Q$:
\begin{equation}
 Q = - \frac{1}{32\pi} \int_{M} d^{4}x
 \epsilon_{\mu\nu\rho\sigma}tr[F_{\mu\nu}(x)F_{\rho\sigma}(x)] \in Z
\end{equation}
The topological charge density $q(x)$ is again a perfect derivative
\begin{equation}
 q(x) = - \frac{1}{32\pi}
 \epsilon_{\mu\nu\rho\sigma}tr[F_{\mu\nu}(x)F_{\rho\sigma}(x)]
        =  \partial_{\mu}K_{\mu}(x)
\end{equation}
where the Chern-Simons density $K_{\mu}(x)$ is
\begin{equation}
 K_{\mu}(x) = - \frac{1}{8\pi^{2}} \epsilon_{\mu\nu\rho\sigma}
   tr[A_{\nu}(x)(\partial_{\rho}A_{\sigma}(x)+
   \frac{2}{3}A_{\rho}(x)A_{\sigma}(x))].
\end{equation}
It transforms like:
\begin{eqnarray}
 K_{\mu}(x) \rightarrow K_{\mu}(x) & - & \frac{1}{24\pi^2}
  \epsilon_{\mu\nu\rho\sigma}tr[g^{-1}(x)\partial_{\nu}g(x)
 g^{-1}(x)\partial_{\rho}g(x)g^{-1}(x)\partial_{\sigma}g(x)] \nonumber \\
                                   & - & \frac{1}{8\pi^2}
  \epsilon_{\mu\nu\rho\sigma}tr[\partial_{\nu}(\partial_{\rho}g(x)
 g^{-1}(x)A_{\sigma}(x))].
\end{eqnarray}
Finally           the Chern-Simons  number $N_{CS}$ is:
\begin{equation}
 N_{CS} = \int_{\partial M=S^3} d^{3}x K_{0}(x) \not \in Z
\end{equation}
This time  the boundary of the manifold $\partial M$
is assumed to be a three sphere $S^3$.
While $N_{CS}$ is only an integer for pure gauge configurations,
the gauge variation is an integer (the boundary term vanishes)
\begin{equation}
 \delta N_{CS} = - \frac{1}{24\pi^2}
          \int_{\partial M=S^3} d^{3}x
  \epsilon_{0\nu\rho\sigma}tr[g^{-1}(x)\partial_{\nu}g(x)
 g^{-1}(x)\partial_{\rho}g(x)g^{-1}(x)\partial_{\sigma}g(x)] \in Z
\end{equation}
This follows also from homotopy theory using the mapping
\begin{equation}
  g(x): S^{3} \rightarrow SU(2) = S^{3}.
\end{equation}
Such mappings are characterized with the homotopy class
\begin{equation}
\Pi_{3}(S^{3}) \in Z.
\end{equation}

\section{Topological charge and the Chern-Simons term on the lattice}
We will now consider the lattice versions of the topological charge
and the  Chern-Simons number. We begin with the two dimensional U(1)
theory and we assume that the manifold is a two torus
 $M = T^2$ and we shall cover $M$ with cells (plaquettes) $c(n)$.
Let the  gauge potential
$A_{\nu}^{n-\hat{\mu}}$ be defined on $c(n-\hat{\mu})$ and likewise
$A_{\nu}^{n}$ be defined on $c(n)$. At the faces  (links)
$f(n,\mu) = c(n-\hat{\mu}) \cap c(n)$, we relate
the two potentials via  a
gauge transformation or transition function $v_{n,\mu}$
\begin{equation}
  A_{\nu}^{n-\hat{\mu}}  = A_{\nu}^{n}  + i
  v^{-1}_{n,\mu}\partial_{\nu}v_{n,\mu}.
\end{equation}
Starting from eqn. 2.3 we find  for the topological charge
\begin{equation}
Q = \frac{i}{2\pi} \sum_{n,\mu} \epsilon_{\mu\nu} \int_{f(n,\mu)}dx_{\nu}
  v^{-1}_{n,\mu}\partial_{\nu}v_{n,\mu} \in Z.
\end{equation}

In the local axial gauge in $c(n)$ one has at the corners
\begin{equation}
 v_{n,\mu}(x) = w_{n-\hat{\mu}}(x)w^{-1}_{n}(x)
\end{equation}
where $w_{n}(x)$ is a parallel transporter.
To get to the lattice result one must interpolate $v_{n,\mu}(x)$
continuosly between $v_{n,\mu}(n)$ and $v_{n,\mu}(n+\hat{\nu})$ say
\begin{equation}
 v_{n,\mu}(x)  =  [v_{n,\mu}(n+\hat{\nu})v^{-1}_{n,\mu}(n)]^{x}
 v_{n,\mu}(n)
                                \;\; 0 \leq x \leq 1.
\end{equation}
After some simple algebra one arrives at:
\begin{equation}
Q = \frac{1}{2\pi} \sum_{n} arg[U_{n,0}U_{n+\hat{0},1}
                 U_{n+\hat{1},0}^{-1}U_{n,1}^{-1}] \in Z,
                    \;\; - \pi < arg < \pi.
\end{equation}
Here $U_{n,\mu}$ are the links and we notice that the charge only depends
on the plaquette angle, hence it is gauge invariant. If we write
$U_{n,\mu} = exp[ik_{n,\mu}]$ then the natural definition of
Chern-Simons term is
\begin{equation}
   N_{CS} = \frac{1}{2\pi} \sum_{n} k_{n,0}
\end{equation}
where the sum is over the spatial lattice only.
Since under a gauge transformation $g(n) = exp[i\theta(n)]$
\begin{equation}
   k_{n,0}  \rightarrow k_{n,0} +
    \theta(n) - \theta(n+\hat{0}) + 2\pi \times m, \;\;  m \in Z,
\end{equation}
it is clear that $N_{CS}$ only can change by an integer.

This generalizes to SU(2) in four dimensions $M = T^4$. The cells are
now hypercubes and the faces $f(n,\mu)$ are cubes. There are also
plaquettes $p(n,\mu,\nu) = f(n) \cap f(n-\hat{\nu})$.
For Q one has, see ref.~\cite{Luesch}.
\begin{eqnarray}
 Q    = & - & \frac{1}{24\pi^2} \sum_{n} [\epsilon_{\mu\nu\rho\sigma}
        \nonumber \\
        &   & \int_{f(n,\mu)} d^{3}
  tr[v^{-1}_{n,\mu}(x)\partial_{\nu}v_{n,\mu}(x)
     v^{-1}_{n,\mu}(x)\partial_{\rho}v_{n,\mu}(x)
     v^{-1}_{n,\mu}(x)\partial_{\sigma}v_{n,\mu}(x)] \nonumber \\
        & + 3 & \int_{p(n,\mu,\nu)} d^{2}x
  tr[v_{n,\mu}(x)\partial_{\rho}v^{-1}_{n,\mu}(x)
   v^{-1}_{n-\hat{\mu},\nu}(x)\partial_{\sigma}v_{n-\hat{\mu},\nu}(x)]].
\end{eqnarray}
At this point we shall deviate from Lueschers charge and use
a slightly different procedure given by Seiberg.
It has the advantage that one can define a Chern-Simons term also.
The topological charge is:
\begin{eqnarray}
Q  & = & \sum_{n} q(n) \nonumber \\
q(n) & = & \frac{1}{2\pi}
             \sum_{\mu}(-1)^{\mu}(k_{n,\mu} - k_{n+\mu,\mu})
                    \;\; - \pi < q(n) < \pi.
\end{eqnarray}
Here $k_{n,\mu}$ is the local Chern-Simons term which leads to
\begin{equation}
   N_{CS}   =   \frac{1}{2\pi} \sum_{n} k_{n,0}
\end{equation}
with the summation over the spatial lattice only. Explicitly:
\begin{eqnarray}
 k_{n,\mu} = & - & \frac{1}{12\pi} \sum_{n}  \epsilon_{\mu\nu\rho\sigma}
        \nonumber \\
        &   & [ \int_{f(n,\mu)} d^{3}
  tr[S(x)\partial_{\nu}S(x)^{-1}
     S(x)\partial_{\rho}S(x)^{-1}
     S(x)\partial_{\sigma}S(x)^{-1}] \nonumber \\
        & + 3 & \int_{p(n,\mu,\nu)} d^{2}x
  tr[P^{-1}(x)\partial_{\rho}P(x)
     S^{-1}(x)\partial_{\sigma}S(x)]  ].
\end{eqnarray}
 The function $P$ is defined on a plaquette with corners $a,b,c,d$:
\begin{equation}
 P(x,y) =
 U^{y}_{ac}[U^{y}_{ca}(U_{ac}U_{cd}U_{db}U_{ba})^{y}
 U_{ab}U^{y}_{bd}]^{x}, \;\;  0\leq (x,y) \leq 1.
\end{equation}
In particular at the corners of the plaquette:
\begin{eqnarray}
 P(0,0) & = &  1 \nonumber \\
 P(1,0) & = &  U_{ab} \nonumber \\
 P(0,1) & = &  U_{ac}  \nonumber \\
 P(1,1) & = &  U_{ac}U_{cd}.
\end{eqnarray}
 Therefore, $P$ can be interpreted as the gauge transformation on the
plaquette which brings the links into the complete axial gauge.
 Likewise $S$ is defined on the cube, such that it brings the
links into the complete axial gauge. We have not given the
expressions for $S$ since they are rather cumbersome. Letting
$U_{n,\mu} = exp[aA_{n,\mu}]$, and taking the naive continuum limit
$a \rightarrow 0$ one finds:
\begin{equation}
 \frac{1}{2\pi} k_{n,\mu}(x)
    = - \frac{a^3}{8\pi^2} \epsilon_{\mu\nu\rho\sigma}
   tr[A_{n,\nu}(\partial_{\rho}A_{n,\sigma}+
   \frac{2}{3}A_{n,\rho}A_{n,\sigma})]
\end{equation}
which agrees with the continuum expression eqn. 2.16.
The only difference between  Luescher's and Seiberg's definition is that
Luescher gauge fixes to the complete axial gauge in the whole hypercube.
All links in the hypercube are therefore typically of the form
$U_{ac}U_{cd}U_{db}U_{ba}$. Thus
the $k_{n,\mu}$ is by itself gauge invariant
and it contributes $O(a^{4})$ in the naive continuum limit. So it
can not be interpreted as a Chern-Simons term.
Also         one can relax the restriction of the topological charge
given in eqn. 3.8.

\section{Tests and Monte-Carlo results for the Chern-Simons density}

We have used an SU(2) Higgs model with action:
\begin{equation}
  S = - \frac{\beta}{2} \sum_{n,\mu <\nu}tr[U_{n,\mu\nu}]
      - \kappa \sum_{\mu}tr[\Phi^{\dagger}_{n}U_{n,\mu}\Phi_{n+\mu}]
      + \lambda (\Phi^{\dagger}_{n}\Phi_{n} - 1)^2.
\end{equation}
We always tried to work close to the Higgs phase transition, so the
gauge Higgs couplings were chosen with:
$(\beta,\lambda) = (2.25,0.5)$ and  $\kappa=0.25,0.30$.

To do the integrals in eqn. 3.10 we have used a vectorized version of
the topology program used in
ref.~\cite{Fox}. One integration is done in analytic
form and we are left with a two dimensional integral.
We have used  the following strategy, which turned out to be the most
efficient. Perform a Gaussian integration with $8\times 8$ points and
store the results for the eight $k_{n,\mu}$ in each cube.
Redo the same thing with $16\times 16$ points and compare the results
for each $k_{n,\mu}$. If the difference is less than 0.001 we accept
the contribution. Otherwise we collect the cubes who's integrals
have not yet converged and redo these with $32\times 32$ points instead.
Compare with the previous values. Usually, at
this point only a few integrals have not converged. Those are normally
quite tricky, so for these we use a library integration
routine with interval adaption. The typically time for one topological
charge on a $6^4$ lattice is 100 seconds on the CRAY-YMP. The charges are
integers up to errors of the order $10^{-4}$. To make the Seiberg
charge converge, it is necessary to perform a global
Landau gauge fixing before the integration.  This is allowed since
the charge is gauge invariant. We have checked that
Lueschers and Seibergs charge definitions agree in each hypercube
up to an integer. Atmost      a few    of hypercubes have a charge
outside the interval $]-1/2,1/2[$, so that often the two definitions
agree.
Now for the Chern-Simons term we only need to evaluate $k_{n,0}$ for
one timeslice. That takes around 5 seconds.
As we know the Chern-Simons term is gauge dependent, but for a  check we
looked at the periodic structure for a $4^3\times 2$ lattice.
We did 100 configurations and we first measured $N_{CS}$ without any
gauge fixing. This required a much higher accuracy for our integrals
before convergence.
In Fig.~1 we have plotted the Chern-Simons
density. Notice that    there are many configurations which have
$N_{CS}$ close to an integer. These configurations can be
interpreted as being  pure gauge.
In Fig.~2 we have plotted the same quantity but with one Landau
gauge fixing sweep. There are now more configurations with
$N_{CS}$ centered around $0,\pm 1$.
Most of the configurations have
changed Chern-Simons number by an integer.
In Fig.~3 we have performed 5 Landau gauge fixing sweeps and
$N_{CS}$   is now centered at  0.
Since we have demonstrated that the Chern-Simons
term only changes by an integer  under gauge transformations, we can
safely restrict $N_{CS}$ to the interval ]-1/2,1/2[. We now compared
the density for the two lattices $6^4$ and $6^{3}\times 2$.
For these and all the other lattices we used 50 Landau gauge fixing
sweeps, this is sufficient for the integrals to converge.
Typically we have around 1000 Chern-Simons numbers.
See Fig.~4 and Fig.~5. There is a trend in the direction of a
flatter distribution at finite temperature. We would like to interprete
this as the system likes to tunnel more often.
We have also done quite a number of other symmetric
lattices $4^{4},6^{4}$ and $8^{4}$,
as well as
asymmetric $8^{3} \times 2,4$.
We emphasize that these results are only
preliminary and a lot has to be done before we can really say that
tunneling is improved at high temperature. We also need to see how  the
density depends on the spatial volume.
Finally we would like to mention that an alternative definition of
the Chern-Simons term has been derived by the authors in
ref.~\cite{Gock}.
\newpage

\newpage
\section{Figure Caption}
Figure 1.
The Chern-Simons density as a function of
the Chern-Simons number
$ N_{CS}$ with zero gauge fixing sweeps.
The volume is denoted $V$, and
$\beta,\kappa,\lambda$ are the gauge Higgs couplings.
Figure 3.
The Chern-Simons density as a function of the Chern-Simons number
$N_{CS}$ with one Landau gauge fixing sweep.
Same parameters as before.
Figure 3.
The Chern-Simons density as a function of the Chern-Simons number
$N_{CS}$ with five Landau gauge fixing sweeps.
Same parameters as before.
Figure 3.
The Chern-Simons density at zero temperature
as a function of the restricted Chern-Simons number
$N_{CS}$ with fifty Landau gauge fixing sweeps.
Same parameters as before.
Figure 5.
The Chern-Simons density at finite temperature
as a function of the restricted Chern-Simons number
$N_{CS}$ with fifty Landau gauge fixing sweeps.
Same parameters as before.
\end{document}